\documentclass[twocolumn,showpacs,preprintnumbers,pre]{revtex4}

	\usepackage[utf8]{inputenc} % To use Unicode characters
	\usepackage[iso]{datetime}
	\usepackage{amsmath, amssymb}
	\usepackage[a4paper]{geometry}
	\usepackage{xcolor}
	\usepackage[colorlinks=true,linkcolor=red,urlcolor=blue,citecolor=red]%
		{hyperref}
	\usepackage{graphicx,epstopdf}
        \usepackage{multirow}
	\usepackage[caption=false]{subfig}
	\newcommand{\reffig}[1]{Fig.~\ref{#1}}
	\newcommand{\reftbl}[1]{Table~\ref{#1}}
	\usepackage{amsmath, amssymb,amsfonts,amsthm}	
	\newcommand{\hbAbs}[1]{| #1 |}	% absolute value
	\newcommand{\argmax}[2]{\underset{#1}{\operatorname{arg \, max}}\;#2}	
	\theoremstyle{plain}% default
	\theoremstyle{definition}
	\usepackage{changes}	
	\definechangesauthor[name=Haluk Bingol,color=purple]{hb}
	\definechangesauthor[name=Mursel Tasgin,color=blue]{mt}
	\setremarkmarkup{(#2)}

\usepackage{float}
\usepackage{array}
\usepackage{booktabs}
\usepackage{xspace} 
\usepackage{textgreek}

\usepackage{clrscode3e} % algorithm pack

\newcommand{\SoL}{\mathbb{L}}	% set of labels 

\newcommand{\BIJ}{b_{i}(j)}

\newcommand{\IR}{I-R}
\newcommand{\ICC}{I-CC}
\newcommand{\ICN}{I-CN}
\newcommand{\GR}{G-R}
\newcommand{\GCC}{G-CC}
\newcommand{\GCN}{G-CN}
\newcommand{\GM}{G-1}
\newcommand{\mtXieSymanski}{LPAc}

\newcommand{\mtCommonNeighbors}[2]{\cap_{#1 #2}}

\newcommand{\aXS}{Xie and Szymanski}
\begin{document}

% =======================================
\title{
	Community detection using boundary nodes in complex networks
}
\author{Mursel Tasgin}
\author{Haluk O. Bingol}
\affiliation{
	Department of Computer Engineering\\
	Bogazici University, Istanbul
}

% =======================================
\begin{abstract}
%-- 
We propose a new local community detection algorithm 
that finds communities by identifying borderlines between them using boundary nodes. 
Our method performs label propagation
for community detection,
where nodes decide
their labels based on the largest ``benefit score'' 
exhibited by their immediate neighbors as an attractor to their communities.
We try different metrics and find that 
using the number of common neighbors as benefit scores
leads to better decisions for 
community structure. 
The proposed algorithm has a local approach 
and focuses only on boundary nodes
during iterations of label propagation,
which eliminates 
unnecessary steps and
shortens the overall execution time.
It preserves small communities
as well as big ones and can outperform other algorithms in terms of 
the quality of the identified communities,
especially when the community structure is subtle.
The algorithm has a distributed nature 
and can be used on large networks in a parallel fashion. 
\end{abstract}

\pacs{89.75.Hc, 89.65.Ef, 89.75.Fb}
\maketitle

% =======================================
% =======================================
\section{Introduction}

%--
A system consisting of elements can be expressed
by using network representation,
i.e.,
nodes denote the elements and 
edges represent their relations. 
Many real-life systems,
e.g., 
mobile communication networks, 
collaboration networks,
protein-protein interaction networks are analyzed using 
network representation~\cite{%
	onnela2007PNAS,
	newman2001Collaboration,
	chen2006Protein}.
A \emph{community} is defined as a group of nodes in a network 
where nodes within the same group have more connections with 
each other than 
the nodes from other groups~\cite{%
						    girvan2002community}. 
\emph{Community detection} is the task of identifying such groups in a network.
Although there is not a universally accepted definition of a community,
the above definition is used by many community detection algorithms~\cite{%
	girvan2002community,
	newman2004fast,
	clauset2004heap,
	rosvall2007Infomap,
	blondel2008Louvain,
	lancichinetti2011OSLOM,   
	de2014mixing,  
	raghavan2007LPA,
	xie2011community,
	gregory2010COPRA,
	eustace2015community,
	tasgin2018preference}.
There is a comprehensive survey on community detection methods 
and algorithms in complex networks
by Fortunato~\cite{%
	fortunato2010Survey}.
Different aspects and purposes of community detection 
are investigated in a recent 
work by Schaub et al.\cite{%
	Schaub2017}.
Authors discuss that understanding the motivation of community detection 
for a specific problem is important for 
selecting the most suitable algorithm or approach,
since there are many facets of community detection.

Many of the proposed community detection algorithms, 
some of which are nearly a decade old or more, 
are successful on small networks of hundreds or thousands of nodes.
With the availability of very large network datasets 
having millions or billions of nodes and edges
in recent years,
there are challenges for community detection algorithms.
Many of the existing community detection algorithms are not able to
run on such large networks
because of their high time-complexity.
If a community detection algorithm needs to optimize a global value or 
a metric regarding the whole network,
then it may need to perform an operation or 
calculation related 
with all elements of the network (i.e. nodes and edges) many times.
Such an approach is computationally expensive and is not feasible
on very large networks.
Additionally,
processing the whole network data may require storing 
and accessing it many times, which is expensive in 
terms of data storage, too.
A \emph{local community detection} approach,
which uses local information around a node while identifying its community,
can be a practical solution on very large networks.
When the community of each node is decided 
using such a limited data and 
calculation, 
then overall time-complexity of the algorithm will be reasonably low
on very large networks.
Besides their practicality, local algorithms may be 
the only viable options 
on these networks.

In this paper, we propose a new community detection algorithm 
that has a local approach 
and tries to find communities by identifying borderlines between them 
using boundary nodes.
Initially,
every node is considered to be a boundary node.
Our community detection process naturally 
decreases their numbers by identifying communities of them.
In the final situation,
only the actual boundary nodes remain
and they constitute the borderlines between communities.

Outline of the paper is as follows.
We first give background information about our notation,
local algorithms and our method of testing.
Then we briefly explain our community detection approach.
We go into the details of experiments and 
present the results of our algorithm on 
both generated and real-life networks
and compare it with other algorithms.

% =======================================
\section{Background}

% =======================================
\subsection{Notation}
\label{sec:notation}

Let $G = (V, E)$ be an unweighted and undirected graph 
where 
$V$ is the set of nodes and 
$E$ is the set of edges.
A \emph{community structure} is a partition of $V$.
We label each block in the partition 
using a symbol in the set of \emph{community labels}
$\SoL = \{ 1, \dotsc, \hbAbs{V} \}$.
We define function
$L \colon V \to \SoL$, 
which maps each node in $V$ to a community label in $\SoL$.
That is, 
the community of node $i \in V$ is given as $L(i)$.
If two nodes $i$ and $j$ are in the same community,
then we have $L(i) = L(j)$.

In community detection,
\emph{triangles},
i.e.,
three nodes connected by three edges,
play an important role~\cite{%
	radicchi2004Clustering}.
We use two metrics related to triangles.
First one, the \emph{clustering coefficient} $CC_{i}$ of node $i$, is 
the probability that two of its neighbors are friends of each other,
given as
\[
	CC_{i}= \frac{\bigtriangleup_{i}}{\wedge_{i}}
\]
where 
$\bigtriangleup_{i}$ is the number of triangles around node $i$ and 
$\wedge_{i}$ is the number of \emph{triplets}, 
i.e.,
$i$ is connected to two nodes,
centered at 
$i$~\cite{%
	newman2001clustering}.
The second metric is the number of common neighbors of two nodes, 
which is generally used for node similarity.
The \emph{number of common neighbors} of nodes $i$ and $j$ is given as
\[
	\mtCommonNeighbors{i}{j} = \hbAbs{\Gamma(i) \cap \Gamma(j)}
\]
where
$\Gamma(i)$ is the \emph{1-neighborhood} of $i$,
i.e.,
the set of nodes whose distances to $i$ are 1.

We use the concepts of \aXS~\cite{
	xie2011community} 
to mark the nodes.
A node $i$ is called an \emph{interior node} 
if it is in the same community with all of its 1-neighbors. 
If it is not an interior node, it is called a \emph{boundary node}.
Note that boundary nodes are positioned 
among nodes from different communities.

% =======================================
\subsection{Local community detection algorithms}

In recent years, several local community detection 
algorithms have been proposed~\cite{%
	raghavan2007LPA,
	xie2011community,
	gregory2010COPRA,
	eustace2015community,
	tasgin2018preference}.
These algorithms generally discover 
communities using local interactions of nodes 
or local metrics calculated in the 1-neighborhood of nodes in the network.
Instead of performing a search or a 
calculation on the whole network (i.e. global), 
local approach splits the community detection task 
into separate subtasks
on individual nodes and their neighborhoods.
Results of these subtasks are then
merged together to get the 
community structure of the whole network.

Raghavan et al.~\cite{%
	raghavan2007LPA} 
proposed label propagation algorithm,
denoted by \emph{LPA}, 
which updates the community label of each node with
the most popular label in its 1-neighborhood,
i.e., 
majority rule of labels.
Labels of all nodes in the network are updated asynchronously and 
algorithm terminates 
when there is no possible label update in the network.
It is a linear-time algorithm, 
which can identify communities in a fast way.
However,
it tends to find a single large community,
especially when community structure is subtle.

\aXS~\cite{%
	xie2011community} 
proposed an extension on LPA, 
which we denote by \emph{\mtXieSymanski}, 
using neighborhood-strength driven approach.
\mtXieSymanski\ improves the quality of identified communities
by incorporating the number of common neighbors
to the majority rule of labels in LPA.
It calculates the scores of labels by 
first counting the number of members 
having these labels, 
which is similar to LPA.
Then it adds the number of common 
neighbors each group has with the node,
multiplied with a constant, $c < 1$.

Additionally,
LPAc also decreases the number of execution steps by 
avoiding unnecessary label updates.
Only a subset of nodes in the network update their labels,
namely, active boundary nodes.
Algorithm defines a node as \emph{passive} if it
would not change its label when there is 
an attempt to update it;
a node that is not passive is called \emph{active}.
It keeps a list of both types and
iteratively selects a node $i$ from the 
active boundary list and updates its label, $L(i)$.
After the label update, 
status of node $i$ is checked and 
if it becomes a passive or an interior node,
it is removed from active boundary list.

After label update, neighbors of $i$ are checked for a change of status,
i.e.,
if they become active boundary nodes,
they are inserted into the list;
if they change from active to passive, 
they are removed from the list.
The algorithm iteratively identifies 
the labels of nodes in active boundary list and
maintains the list with removals and 
insertions of nodes with label updates.
Algorithm completes when the active boundary list is empty.
Despite the increased quality of communities and its speed,
LPAc still has the issue of finding a single 
community, which is a drawback of LPA, too.

% =======================================
\subsection{Method for testing of algorithms}

A community detection algorithm outputs a partition of the set of vertices,
where each block of the partition corresponds to a community.
When we have the ground-truth community structure of the network,
we can compare the partition output of 
the algorithm with that of the ground-truth
using Normalized Mutual Information (\emph{NMI})~\cite{%
	danon2005NMI}.

We start testing our algorithm on real-life 
networks with ground-truth community structure.
The first network is the small network of Zachary karate club~\cite{%
	zachary1977Karate}.
Then, we use larger networks provided by 
SNAP~\cite{%
	datasetSNAP2014}, 
namely;
DBLP network, 
Amazon co-purchase network, YouTube network and 
European-email network,
which all have ground-truth communities.
Although we use the provided
ground-truth communities,
which are created by using some meta-data related to these networks,
Peel et al.~\cite{%
             Peel2017}
present a detailed analysis on whether the given meta-data can 
explain the actual ground-truth communities for the corresponding network.
When an algorithm finds the communities
on a network that are different from the communities
explained by meta-data, 
then it may not be directly related with algorithm's failure;
but there may be other reasons, 
i.e. 
irrelevant meta-data, 
meta-data showing different aspects of the network or 
no community structure in the network.

On real-life networks, 
we run some of the known community detection algorithms;
namely 
Newman's fast greedy algorithm 
(\emph{NM})~\cite{%
	clauset2004heap}, 
Infomap 
(\emph{Inf})~\cite{%
	rosvall2007Infomap}, 
Louvain 
(\emph{Lvn})~\cite{%
	blondel2008Louvain},
LPA~\cite{%
	raghavan2007LPA},
and
LPAc~\cite{%
	xie2011community} 
and compare their results with the ground-truth. 
Execution times of the algorithms are also measured and reported.
Experiments are done with a computer 
having 2.2 GHz Intel Core i7 processor with 4-cores.

We also use a set of computer generated networks for testing.
The LFR benchmark networks~\cite{%
	lancichinetti2008benchmark}, 
with planted community structure, 
are used for comparison of community detection algorithms.
These networks are generated with a parameter vector of 
$[N, \langle k\rangle, k_{max}, C_{min}, C_{max}, \mu]$
where 
$N$ is the number of nodes and 
$\mu$ is the mixing parameter controlling the rate of 
intra-community edges to all edges of nodes in the generated network.
Community structure of an LFR network is 
related to the mixing parameter it is generated with.
As $\mu$ increases, 
the community structure becomes more subtle and difficult to identify.
LFR algorithm runs in a non-deterministic way and 
can create different networks, 
given the same parameter vector.
In order to avoid a potential bias of an algorithm to a single network,
we generate 100 LFR networks for each vector and report the averages.

% =======================================
\section{Our Approach}

We propose a new community detection algorithm that finds communities
by identifying borderlines between communities based on boundary nodes.
We first provide an overview of the algorithm,
then we discuss the details.
The algorithm $\proc{Community-By-BoundryNodes}$ 
is given in \reffig{fig:Community-By-BoundryNodes}.

% +++++++++++++++++++++++++++++++++++++++
%: ++ alg:Community-By-BoundryNodes
\begin{figure}[thbp]
\begin{codebox}
	\Procname{$\proc{Community-By-BoundryNodes}$}
	\li \Comment  $V \gets \{ 1, \cdots, |V| \}$: set of nodes
	\li \Comment  $S$: set of boundary nodes
	\li \Comment  $L[i]$: the community of $i \in V$
	\li \Comment $\proc{Initial-Heuristic}()$: an initial heuristic
	\li \Comment  $\proc{isBoundryNode}(i)$: $\const{true}$ 
	\li \Comment  $          $     if $i \in V$ is a boundary node
	\li \Comment  $\proc{bestCommunity}(i)$: best community 
	\li \Comment  $          $     for $i \in V$
	\li
	% ====
	\li \Comment initialization
	\li \While $i$ in $V$
	\li \Do
		$L[i] \gets i$
	\End
	\li $\proc{Initial-Heuristic}()$
	% ====
	\li \While $i$ in $V$
          \Do
          \li \If $\proc{isBoundryNode}(i)$
           \Do
	   \li $S \gets S \cup \{i\}$
	   \End
	\End
	\li
	% ====
	\li \Comment iteration	
	\li \While $S \ne \emptyset$
	\li \Do
		    $i \gets$ randomly selected node in $S$
		\li $S \gets S \smallsetminus \{ i \}$
		\li $communityOld \gets L[i]$
		\li $ L[i] \gets \proc{bestCommunity}(i)$
		\li \If $communityOld \ne  L[i]$
		\li \Do
			    \While $j$ in $\Gamma(i)$
			\li \Do
				    \If $\proc{isBoundryNode}(j)$
				\Do
					\li $S \gets S \cup \{ j \} $
				\End
			\End
		\End
	\End
\end{codebox}
\caption{
	Main algorithm \proc{Community-By-BoundryNodes}.
}
\label{fig:Community-By-BoundryNodes}
\end{figure}
% +++++++++++++++++++++++++++++++++++++++

The algorithm keeps track of a set $S$ of boundary nodes.
We start with $|V|$ communities of size 1,
i.e., 
each node is a community by itself.
Since each node is a boundary node,
the set initially 
would have all the nodes in it.

Set $S$ with $|V|$-elements is too large.
We apply a heuristic, 
\proc{Initial-Heuristic} in \reffig{fig:Initial-Heuristic},
to reduce the initial number of communities, 
hence, 
the initial number of boundary nodes.
For each connected pair of nodes $i,j\in V$,
we calculate the ``benefit score'', 
$\BIJ$,
if $i$ assumes the community of $j \in \Gamma(i)$.
Note that $\BIJ$ is calculated synchronously.
We set the community of $i$ to that of $j$ with the maximum benefit score.
Then, 
using procedure $\proc{isBoundryNode}$ in \reffig{fig:isBoundryNode},
we identify the boundary nodes and insert them into the set $S$.

% +++++++++++++++++++++++++++++++++++++++
%: ++ alg:Initial-Heuristic
\begin{figure}[thbp]
\begin{codebox}
	\Procname{$\proc{Initial-Heuristic}$}
	\li \Comment $\BIJ$: benefit score if $i$ assumes the community of $j$
	\li
	\li \While $i$ in $V$
	    \Do
	        \li $maxBenefit \gets 0$
	        \li $maxNode \gets 0$
	        \li \While $j$ in $\Gamma(i)$
	          \Do
	              \li \If $\BIJ>  maxBenefit$
	                 \Do
	                     \li $maxBenefit \gets \BIJ$
	                     \li $maxNode \gets j$
	                 \End		      
		  \End
		 \li $L[i] \gets L[maxNode]$
	   \End
\end{codebox}
\caption{
	Procedure  \proc{Initial-Heuristic}.
}
\label{fig:Initial-Heuristic}
\end{figure}
% +++++++++++++++++++++++++++++++++++++++

% +++++++++++++++++++++++++++++++++++++++
%: ++ alg:isBoundryNode
\begin{figure}[thbp]
\begin{codebox}
	\Procname{$\proc{isBoundryNode}(i)$}
	\li \While $j$ in $\Gamma(i)$
	\li \Do
		    \If $L[j] \ne L[i]$
		\Do
			\li \Return \const{true}
		\End
	\End
	\li \Return \const{false}
\end{codebox}
\caption{
	Procedure  \proc{isBoundryNode}.
}
\label{fig:isBoundryNode}
\end{figure}
% +++++++++++++++++++++++++++++++++++++++

As long as the set is not empty,
the algorithm repeats the following steps.
A node $i$ in the set is selected at random 
and removed from the set.
	We reconsider the community of the selected node based on its 1-neighborhood.
	A new community assignment, 
	which produces the largest ``benefit score'', 
	is made.
If the old and the new communities of $i$ are the same,
i.e. no effective change,
then we are done with this pass.
If the community of $i$ is changed,
then this may cause some of its 1-neighbors
to become boundary nodes.
In this case, 
the new boundary nodes are inserted into the set.
Note that the selected node is not added to the set during this iteration
even if it is still a boundary node.
It is possible that it may be inserted into the set in some other iteration,
in which one of its 1-neighbors is processed.
Boundary node check is done with procedure $\proc{isBoundryNode}$.

This iteration process terminates 
when the set $S$ becomes empty,
which indicates that the system reaches to a steady state,
where no further change in community assignment is possible
with a larger ``benefit score''.

% =======================================
\subsection{Best Community}
\label{subsection:BestCommunity}
Given a community assignment,
we want to reconsider the community $L(i)$ of a node $i$ by 
investigating options in its 1-neighborhood $\Gamma(i)$.
This is the function of the procedure 
$\proc{bestCommunity}$,
which is described below.
There are two different approaches:

%: ==== Individual approach 
\textbf{a) Individual approach.}
Consider each neighbor $j \in \Gamma(i)$ of $i$ individually.
Switching to the community of $j$ produces a benefit of $\BIJ$.
Therefore, 
$i$ switches to the community of $j$, 
which produces the largest benefit.
That is,
\[
	L(i) 
	= L\left(
		\argmax {j \in \Gamma(i)} 
			\BIJ
	\right).	
\]
If there is more than one community with the maximum benefit,
one of them is selected randomly.
%

% ====
For the value of $\BIJ$,
we consider three metrics:
(i)~\textbf{\emph{\IR}:}
Assign a uniformly random number in a range of $[0, 1]$ to $\BIJ$.
Clearly,
this will not reflect any information regarding the properties of a node or its neighborhood.
(ii)~\textbf{\emph{\ICC}:}
Use the clustering coefficient of $j$ as the benefit score,
i.e.,
$\BIJ = CC_{j}$.
(iii)~\textbf{\emph{\ICN}:}
Use the number of common neighbors of $i$ and $j$,
i.e., 
$\BIJ = \mtCommonNeighbors{i}{j}$.\\

%: ==== Community groups approach 
\textbf{b) Community groups approach.}
We consider the communities represented by the neighbors. 
The neighbors are grouped according to their communities.
We look at the collective contribution of each group.
The community of the group with the largest benefit score 
is selected as the new community of $i$.
That is,
\[
	L(i) 
	= L\left(
		\argmax{k} 
			B_{i}(k)
	\right)
\]
where 
$B_{i}(k)$ is the collective benefit score of 
community $k$ 
in 1-neighborhood of node $i$
and defined as
\[
	B_{i}(k) 
	= \sum_{
			\substack{
				L(j) = k\\
				j \in \Gamma(i)
			}
		}
		\BIJ.
\]

%====

For the value of $\BIJ$,
we consider the three metrics
that we used in the individual approach.
The group versions are denoted by
(iv):~\textbf{\emph{\GR}},
(v):~\textbf{\emph{\GCC}}, and
(vi):~\textbf{\emph{\GCN}}.
In addition to these,
we consider one more measure:
(vii):~\textbf{\emph{\GM}}:
We assign $\BIJ = 1$ to each neighbor $j$.
Note that this is similar to the majority rule of labels in LPA.

% =======================================
\section{Experiments and Discussion}
\label{section:ExperimentsAndDiscussion}

% =======================================
\subsection{Deciding the benefit score}

\begin{figure}[!htbp]
	\centering
	\subfloat[NMI Scores]{
		\label{fig:LFR1000_benefitScores}
		\includegraphics
			[width=0.45\textwidth]
			{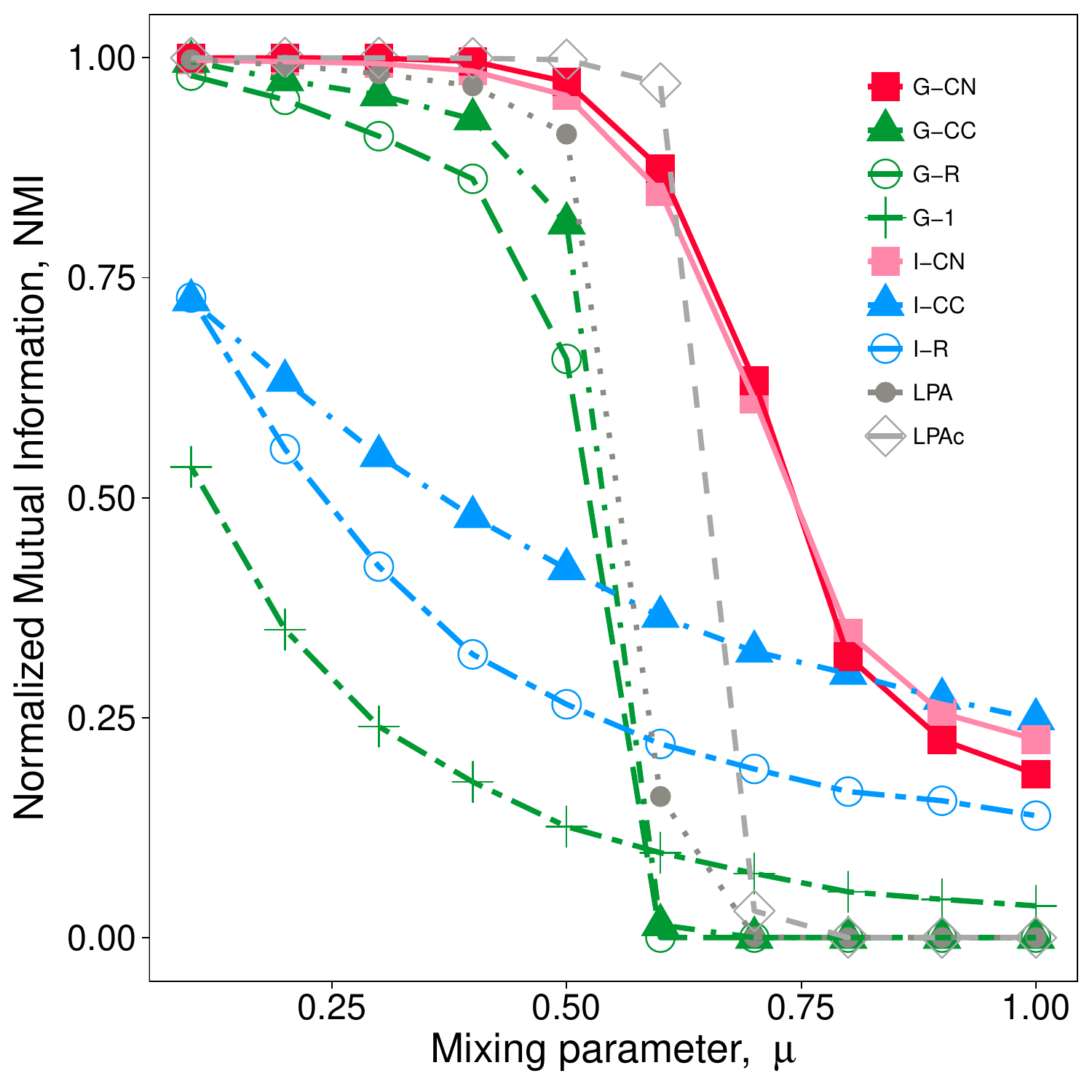}
	}\\
	\subfloat[Execution Times]{
		\label{fig:LFR1000_benefitScoreExecutionTimes}
		\includegraphics
			[width=0.45\textwidth]
			{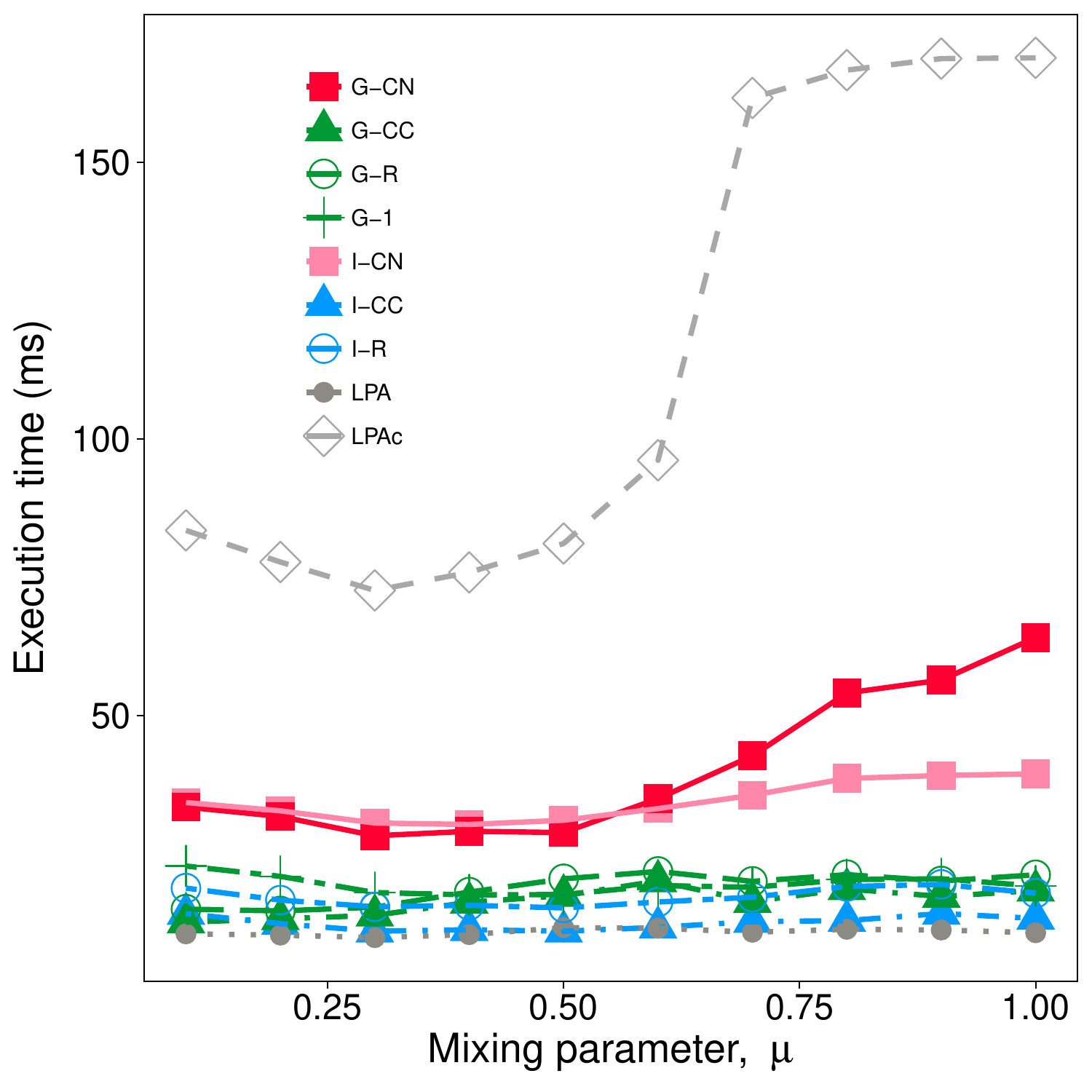}
	}
	\caption{
		\textbf{(a)}~NMI scores and 
		\textbf{(b)}~execution times of our new algorithm 
		with various benefit score candidates, 
		and 
		LPA 
		and LPAc with $c=0.25$
		on LFR benchmark network datasets. 
		LFR parameters:
		$[N=1,000, \langle k\rangle=15, k_{max}=50, C_{min}=10, C_{max}=50]$.
		(Average of 100 realizations)
	}
	\label{fig:LFR1000_Networks_benefitScores}
\end{figure}

We define seven metrics for benefit score.
In order to decide on which metric to use, 
we try each one on LFR generated networks of $1,000$ nodes.
%
%====
NMI scores of identified partitions and execution times of our algorithm are presented 
in \reffig{fig:LFR1000_benefitScores} 
and \reffig{fig:LFR1000_benefitScoreExecutionTimes}, 
respectively.
We also run LPA and \mtXieSymanski\
algorithms on these networks for comparison.
The $c$ parameter of \mtXieSymanski\ is set as $0.25$.

% +++++++++++++++++++++++++++++++++++++++
%: ++ fig:LFR1000_Networks_benefitScores
% +++++++++++++++++++++++++++++++++++++++

We observe that all the group-based benefit scores have 
better results than the individual ones. 
Even group-random value assignment, \GR, has good results.

Surprisingly, 
uniform benefit score using the group approach, \GM,
has the worst performance among all.
Although it is similar to the majority rule of labels in the LPA, 
it is not a good fit for our algorithm.
As an exception among the individual ones,
\ICN\ outperforms LPA.

Benefit scores based on common neighbors,
both at individual and group level,
i.e.,
\ICN\ and \GCN,
produce better results in our tests.
\GCN\ is slightly better than \ICN\ in terms of NMI values.

Both LPA and \mtXieSymanski\ algorithms have good NMI values, 
but when $\mu > 0.6$, 
their performances degrade while our algorithm still finds communities.
\mtXieSymanski\ performs better than LPA.
However, when we look at the execution times of algorithms, 
\mtXieSymanski\ has the worst performance.
Its elapsed time is 2 to 3 times higher than our \GCN\ algorithm.

We conclude that using the number of common neighbors with
the community-groups approach, namely \GCN,
produces the best results in our algorithm.
We use \GCN\ for the rest of the paper.

% =======================================
\subsection{Zachary karate club network}

% +++++++++++++++++++++++++++++++++++++++
%: ++ fig:karateClubCommunities  
\begin{figure}[!htbp]
	\begin{center}
		\includegraphics
			[width=0.7\columnwidth]
			{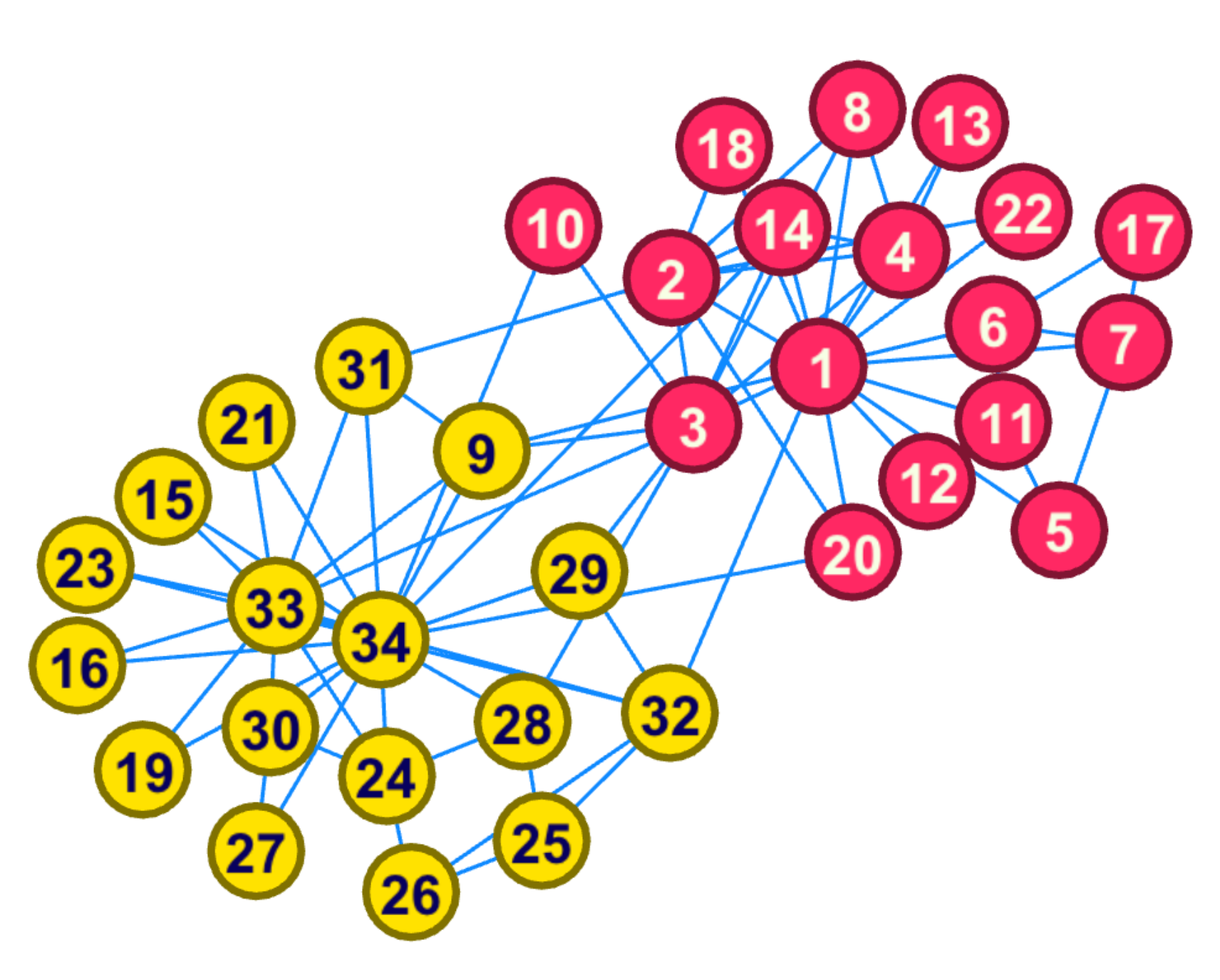}
		\caption{Zachary karate club: Identified communities by our \GCN\ algorithm}
		\label{fig:karateClubCommunities}
	\end{center}
\end{figure} 
% +++++++++++++++++++++++++++++++++++++++

We run our algorithm on Zachary karate club network and 
compare the identified communities with the ground-truth.
Our algorithm \GCN\
identifies two communities as seen in \reffig{fig:karateClubCommunities}.
Only the node~10 is misidentified by our algorithm.
There is a tie-situation among the benefit scores
exhibited to node 10 by its neighbors;
so with random selection among alternatives 
our algorithm sometimes selects the 
\emph{wrong} community.
The community labels of all the other nodes 
are identified correctly with respect to 
ground-truth community structure.

% +++++++++++++++++++++++++++++++++++++++
%: -- table I tbl:tableLargeNetworks
\begin{table*}[!htbp]
	\caption{
		Large real-life networks with ground truth
	}
	\begin{center} 
	\scalebox{0.60}{
		\begin{tabular}{|l|r|r|r|r|r|r|r|r|r|r|r|r|r|r|r|r|r|r|r|r|r|r|}
			% +++++++
			\hline
			\toprule	
			\multirow{2}{*}{Network} &	\multirow{2}{*}{$|V|$} &\multirow{2}{*}{$|E|$}& \multirow{2}{*}{CC} &  \multicolumn{7}{c}{\# communities} & \multicolumn{6}{|c|}{NMI wrt GT}& \multicolumn{6}{c|}{Execution time (ms)}\\ \cline{5-23}
			{} & {} & {} & {} 				& GT		& \GCN  	& Inf 		& LPA &LPAc	&   Lvn  	& NM 	& \GCN	& Inf 	&	LPA	&LPAc&Lvn		&NM	& \GCN	&  Inf 	&	LPA	& LPAc &Lvn		&NM	\\ \hline	                               
European-email	&1,005& 	16,064	&	0.40	&      	  42	&	23		&	  38	&3		&20&	25	&28		&0.14&0.62&0.13&0.31&0.54&0.46 & 146 	 & 133 	& 32 	&704& 69 & 187\\
DBLP	&   	317,080 	&	1,049,866	&	0.63 	& 13,477	&	26,873  	&  30,811	&36,780 	&30,242&	565	&3,165	&0.56&0.65&0.64&0.61&0.13&0.16 & 8,825 	 & 35,753 	& 26,413 &894,858& 8,217 	& 4,362,272\\
Amazon	&	334,863	&	925,872	&	0.40	& 75,149	&	33,395 	&  35,139	&24,045 	&30,908&	248	&1,474	&0.57&0.60&0.54&0.57&0.11&0.11 & 7,552 	 & 43,253 	&30,931&997,088& 8,017 	& 1,422,590\\
YouTube	&	1,134,890	&	2,987,624	&	0.08	&   8,385	&     116,082 	&102,125	&89,449  	&69,817& 9,616&	-	&0.07&0.13&0.07&0.05&0.06&-     & 295,935& 188,037 &324,641&76,129,367& 52,798 	& -\\
			\bottomrule
			\hline
			% +++++++
		\end{tabular}
		}
	\end{center}
	\label{tbl:tableLargeNetworks}
	\scalebox{0.65}{
	\begin{minipage}{0.85\textwidth}%
		\begin{center}
		\begin{tabular}{ll}
			\small GT: Ground-truth
			&\small LPA : Label propagation algorithm~\cite{raghavan2007LPA}\\
			\small \mtXieSymanski: Neighborhood-strength driven LPA~\cite{xie2011community}
			&\small \GCN: Our algorithm\\
			\small Lvn: Louvain community detection algorithm~\cite{blondel2008Louvain}
			&\small Inf: Infomap algorithm~\cite{rosvall2007Infomap}\\
			\small NM: Newman's fast greedy algorithm~\cite{clauset2004heap}\\
		\end{tabular}
		\end{center}
	\end{minipage}%
	}
\end{table*}    
% +++++++++++++++++++++++++++++++++++++++

% =======================================
\subsection{Large real-life networks}

We run our \GCN\  algorithm on large
real-life networks with ground-truth 
communities, provided by SNAP~\cite{
	datasetSNAP2014}. 
For comparative analysis, 
Newman's fast greedy algorithm 
(\emph{NM})~\cite{%
	clauset2004heap}, 
Infomap 
(\emph{Inf})~\cite{%
	rosvall2007Infomap}, 
Louvain 
(\emph{Lvn})~\cite{%
	blondel2008Louvain},
Label Propagation 
(\emph{LPA})~\cite{%
	raghavan2007LPA} and
neighborhood-strength driven LPA
(\emph{\mtXieSymanski})~\cite{xie2011community}
are also run on these networks.
Newman's algorithm is omitted for YouTube network 
due to its long execution time.
The results are presented in \reftbl{tbl:tableLargeNetworks}.

There is no clear winner in \reftbl{tbl:tableLargeNetworks},
which is a good news for local algorithms.
That is, 
although the local algorithms cannot see the global picture, 
they perform good enough.

The number of detected communities by
\GCN, Infomap, LPA, and \mtXieSymanski\ are close to each other
and not far from the ground-truth.
There are exceptions;
on the YouTube network, all four detect too many communities.
On European-email network with 42 ground-truth communities, 
LPA merges many communities together and detects only 3 communities 
while the other three do a better job.
On DBLP and Amazon networks;
both Louvain and 
Newman's algorithm detect very few number of communities.
Louvain has the best detection on YouTube network, 
while Newman's algorithm 
experiences performance problems.
Our \GCN\ algorithm 
performs well on most of the networks. 
However, 
it performs poorly on YouTube network, 
which has the smallest clustering coefficient 
of these four networks.

Considering NMI values of all six algorithms,
it is possible that YouTube network may have subtle community structure.
On this network, the best performing algorithm, Infomap,
only gets NMI value of 0.13.
On DBLP and Amazon networks;
Infomap, LPA, \mtXieSymanski\ and our algorithm obtain similar NMI values, and 
they are much better than Louvain and Newman's algorithm.
On European-email network, 
local algorithms like LPA,
\mtXieSymanski\ and ours are not good enough.
It is possible that the network is not a good one for local approaches.
On all of the networks, \mtXieSymanski\ has highest 
execution times among the local algorithms.
Its execution time on YouTube network is very high compared to our algorithm and LPA.

For all the real-life networks, we use the provided 
ground-truth community structure to evaluate the quality of partitions found by 
each algorithm.
However, there is no single algorithm that performs good on all networks or there is not 
a single network on which some algorithms perform very good.
This may be due to the fact that supposed ground-truth for these networks 
do not reflect the original ground-truth community structure or 
show different aspects of the network structure, 
as discussed in the work of 
Peel et al.~\cite{Peel2017}.
For this reason, our test on real-life networks gives 
an idea about the relative performance of
algorithms compared to each other on different networks;
but does not lead to a conclusion on
whether they perform well on these networks or not.

% =======================================
\subsection{Generated networks}
\label{sec:LFRNetworks}

\begin{figure*}[!htbp]
	\centering
	\subfloat[NMI Scores]{
		\label{fig:LFR1000_NMI}
		\includegraphics
			[width=0.9\columnwidth]
			{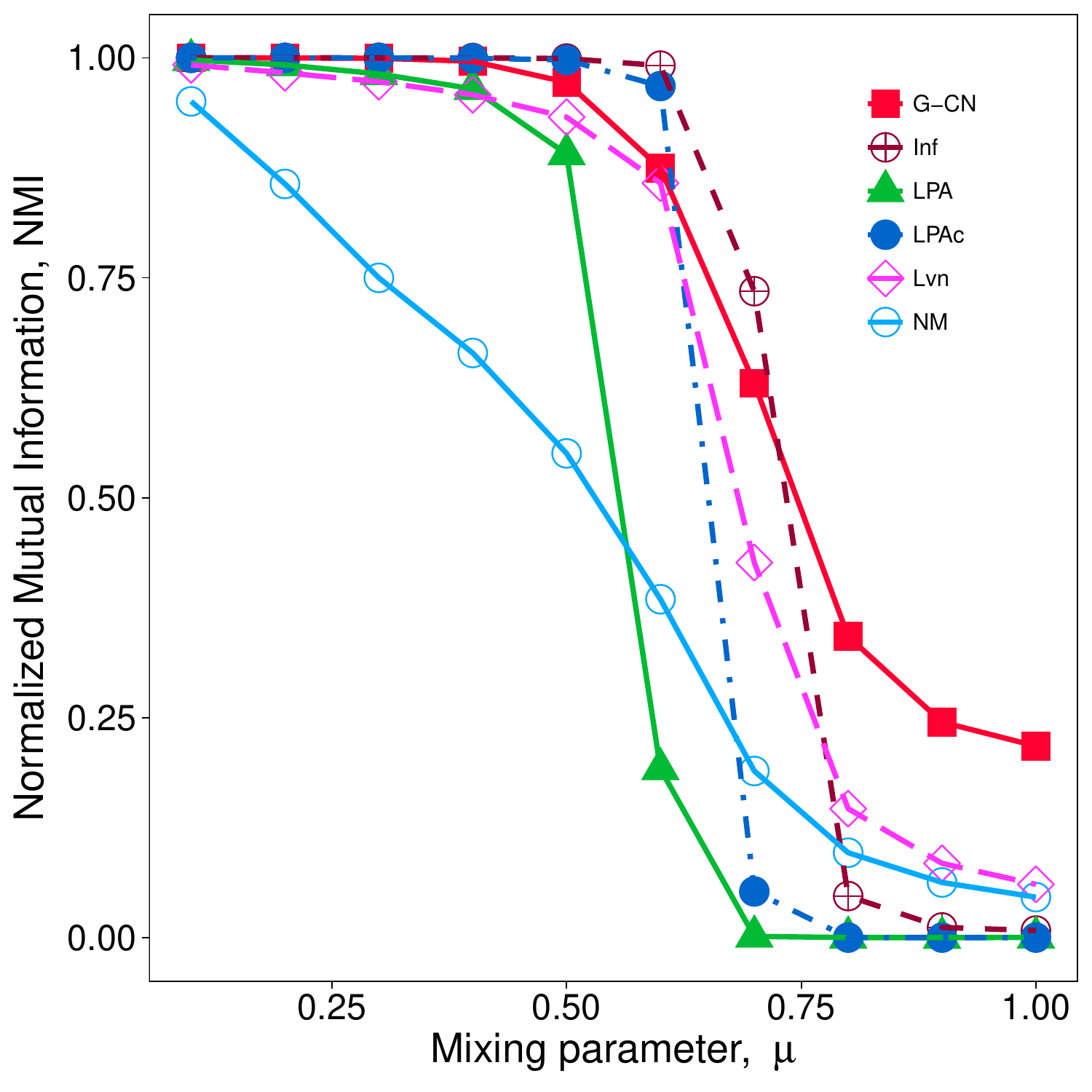}
	}
	\subfloat[Execution Times]{
		\label{fig:LFR1000_executionTimes}
		\includegraphics
			[width=0.9\columnwidth]
			{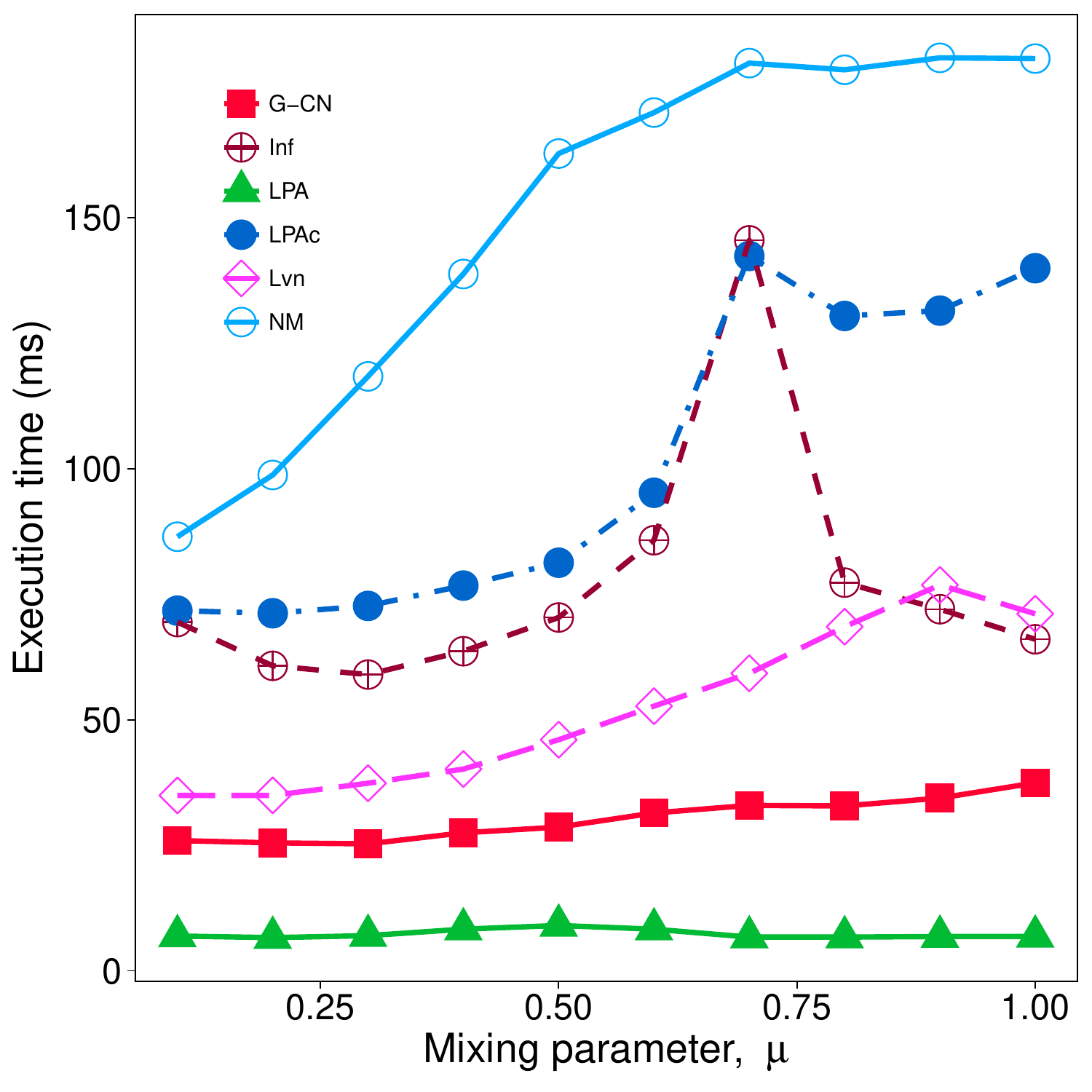}
	}\\
	\subfloat[NMI Scores]{
		\label{fig:LFR5000_NMI}
		\includegraphics
			[width=0.9\columnwidth]
			{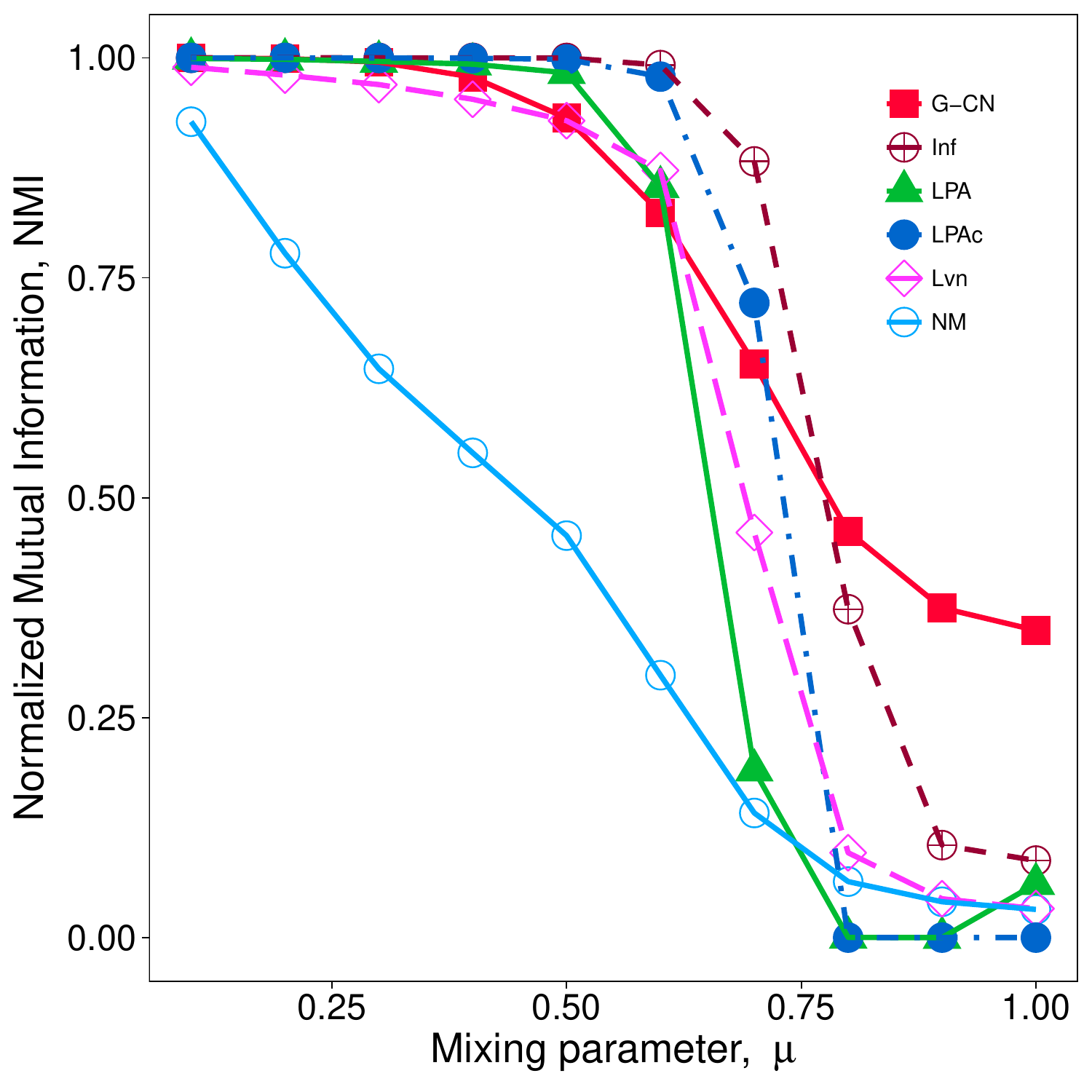}
	}
	\subfloat[Execution Times]{
		\label{fig:LFR5000_executionTimes}
		\includegraphics
			[width=0.9\columnwidth]
			{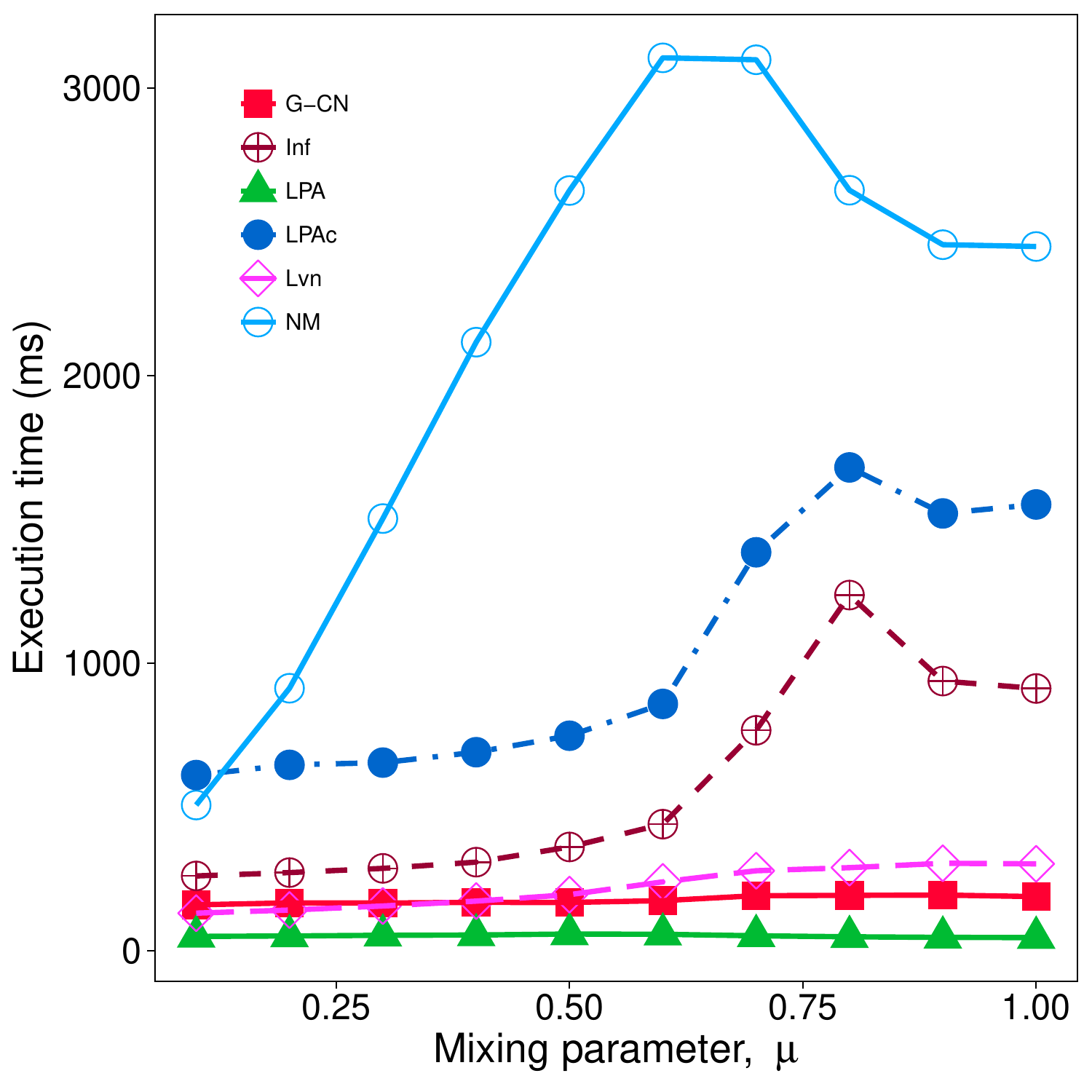}
	}
	\caption{
		Comparison of NMI and execution times of 
		our method and known algorithms 
		on LFR benchmark network datasets. 
		\textbf{(a)} and 
		\textbf{(b)} 
		are for LFR network generated with 
		$[N = 1,000, \langle k\rangle = 15, k_{max} = 50, C_{min} = 10, C_{max} = 50]$.
		\textbf{(c)} and 
		\textbf{(d)} 
		are for LFR network generated with 
		$[N = 5,000, \langle k\rangle = 15, k_{max} = 75, C_{min} = 20, C_{max} = 100]$. (Average of 100 realizations)
	}
	\label{fig:LFR_ComparativeAnalysis}
\end{figure*}

We test our algorithm, \GCN,
also on generated LFR networks 
of $1,000$ and $5,000$ nodes
as reported in \reffig{fig:LFR1000_NMI} and \reffig{fig:LFR5000_NMI}, 
respectively.
The same algorithms 
that we run on real-life networks 
are also used for comparative analysis on these networks.
We also measure the execution times of 
the algorithms and report the results
in \reffig{fig:LFR1000_executionTimes} and 
in \reffig{fig:LFR5000_executionTimes}.
We present the details of the results on LFR 
networks of 5,000 nodes in \reftbl{tbl:tableLFRLarge}.
For each parameter set, 
we generate 100 LFR networks for a given $\mu$ and 
run algorithms on all these datasets and
then average the results for each algorithm.

On LFR networks with 1,000 nodes, our \GCN\ algorithm
is the best algorithm with Infomap when $0.1<\mu<0.5$.
For $0.5<\mu<0.8$, 
our algorithm is in the second place after Infomap.
When $\mu>0.7$, most of the algorithms tend to find 
a small number of communities while our algorithm 
still identifies a reasonable set of communities.
LPA and LPAc find a single community 
that leads to the NMI value of $0$.
Louvain and Newman's 
algorithm also find very few number of communities on these networks.

% +++++++++++++++++++++++++++++++++++++++
%: -- table II  tbl:tableLFRLarge
\begin{table*}[!htbp]
\centering
\caption{Generated LFR benchmark networks of 5,000 nodes  }
\label{tbl:tableLFRLarge}
\scalebox{0.69}{
\begin{tabular}{|l|r|r||r|r|r|r|r|r|r|r|r|r|r|r|r|r|r|r|r|r|r|r|r|}
\hline
\multicolumn{1}{|c|}{\multirow{2}{*}{Network}} & \multicolumn{1}{c|}{\multirow{2}{*}{$|V|$}} & \multicolumn{1}{c|}{\multirow{2}{*}{$\mu$}} & \multicolumn{1}{c|}{\multirow{2}{*}{$|E|$}} 
& \multicolumn{1}{c|}{\multirow{2}{*}{CC}} & \multicolumn{7}{c|}{\# communities}     & \multicolumn{6}{c|}{NMI wrt GT}      & \multicolumn{6}{c|}{Execution time (ms)}       \\ \cline{6-24} 
\multicolumn{1}{|c|}{}                         & \multicolumn{1}{c|}{}                     & \multicolumn{1}{c|}{}                        & \multicolumn{1}{c|}{}                     & \multicolumn{1}{c|}{}                    
& \multicolumn{1}{c|}{GT} & \multicolumn{1}{c|}{\GCN} & \multicolumn{1}{c|}{Inf} & \multicolumn{1}{c|}{LPA} & \multicolumn{1}{c|}{LPAc} & \multicolumn{1}{c|}{Lvn} & \multicolumn{1}{c|}{NM} 
& \multicolumn{1}{c|}{\GCN} & \multicolumn{1}{c|}{Inf} & \multicolumn{1}{c|}{LPA} & \multicolumn{1}{c|}{LPAc}& \multicolumn{1}{c|}{Lvn} & \multicolumn{1}{c|}{NM} 
& \multicolumn{1}{c|}{\GCN} & \multicolumn{1}{c|}{Inf} & \multicolumn{1}{c|}{LPA} & \multicolumn{1}{c|}{LPAc} & \multicolumn{1}{c|}{Lvn} & \multicolumn{1}{c|}{NM} \\ \hline
LFR-1      & 5,000	& 0.1& 38,928& 0.52& 102& 102& 102& 102&102& 89	& 65&  1.00& 1.00 & 1.00 &1.00& 0.99&  0.93 	& 161	& 261	& 51&612&132& 508                     \\ \hline
LFR-2      & 5,000	& 0.2& 38,834& 0.37& 101& 102& 101& 100&101& 81	& 32&  1.00& 1.00 & 1.00 &1.00& 0.98&  0.78 	& 167	&  273	& 52&647&142& 914                      \\ \hline
LFR-3      & 5,000	& 0.3& 38,883& 0.26& 101& 103& 101& 98 &101& 73	& 18&  1.00& 1.00 & 1.00 &1.00& 0.97&  0.65 	&  167	& 287	& 55&655&157& 1,504                    \\ \hline
LFR-4      & 5,000	& 0.4& 38,939& 0.16& 101& 109& 101& 97 &102& 64	& 12&  0.98& 1.00 & 0.99 &1.00& 0.95&  0.55 	&  169	& 309	& 56&691&174& 2,117                    \\ \hline
LFR-5      & 5,000	& 0.5& 38,965& 0.10& 101& 131& 101& 94 &104& 53	&   9&  0.93& 1.00 & 0.98 &1.00& 0.93&  0.46 	&   169	& 362	& 59&749&196& 2,644                    \\ \hline
LFR-6      & 5,000	& 0.6& 38,935& 0.05& 102& 203& 104& 87 &110& 41	& 11&  0.82& 1.00 & 0.85 &0.98& 0.87&  0.30 	&   175	& 441	& 58&859&241& 3,106                    \\ \hline
LFR-7      & 5,000	& 0.7& 38,857& 0.02& 101& 356& 159&  5 &114& 24	& 15&  0.65& 0.88& 0.19 &0.72& 0.46&  0.14 	&   192	& 767	& 53&1,368&279& 3,099                    \\ \hline
LFR-8      & 5,000	& 0.8& 38,873& 0.01& 101& 530& 239&  1 &1& 12 		& 13&  0.46& 0.37& 0.00 &0.00& 0.10&  0.06 	&  193	&  1,238	& 49&1,681&290& 2,645                    \\ \hline
LFR-9     & 5,000	& 0.9& 38,909& 0.01& 102& 614&   86&  1 &1& 12		& 13&  0.37& 0.11& 0.00 &0.00& 0.04&   0.04 	&  195	&  939	& 47&1,522&305& 2,456                    \\ \hline
LFR-10   & 5,000	& 1.0& 38,923& 0.01& 101& 618&   79&  1 &1& 12		& 13&  0.35& 0.09& 0.06 &0.00& 0.03&   0.03 	&  189	&  913	& 47&1,553&303& 2,450                    \\ \hline
\end{tabular}
}
\end{table*}
% +++++++++++++++++++++++++++++++++++++++

% +++++++++++++++++++++++++++++++++++++++
%: ++ fig:LFR_ComparativeAnalysis
% +++++++++++++++++++++++++++++++++++++++

The second set of test is performed on LFR networks of $5,000$ nodes.
Infomap, LPA, and LPAc are successful 
in identifying communities when mixing parameter is low, 
however, their quality degrades with increasing mixing parameter.
LPAc and LPA have slightly better results on
 the networks of $5,000$ nodes generated with $0.4<\mu<0.6$
compared to the previous set of networks of $1,000$ nodes.
With increasing value of the $\mu$, performances of LPA and LPAc get 
worse and they tend to find a single community after $\mu>0.7$.
Infomap has the similar tendency but has better results on 
LFR networks of $5,000$ 
nodes compared to previous 
set of networks of $1,000$ nodes.
Newman's algorithm and Louvain 
find a small number of communities; 
they tend to merge communities, 
which may lead to a resolution limit~\cite{%
	fortunato2007resolution}.
	
Our \GCN\ algorithm identifies communities with 
high accuracy when $\mu$ is low.
It is the only algorithm to identify communities 
when community identification becomes very hard,
i.e.,
$\mu>0.75$.
Its execution times are lower than most of the algorithms;
only LPA has better execution times.
However, considering the quality of identified communities
and corresponding execution times,
\GCN\ algorithm performs better than LPA.
Newman's algorithm and \mtXieSymanski\ have 
the highest execution times on these networks.

% =======================================
\section{Conclusion}

We propose a new local community detection algorithm, \GCN,
which is based on identifying borderlines of communities using boundary nodes in the network.
It is a local algorithm that is able to run on very large 
networks with low execution times.
It can identify communities with high quality, 
regardless of the network size.

On the networks with subtle community structure,
it outperforms other algorithms.
On these networks;
Infomap, LPA, and \mtXieSymanski\ merge all the nodes into a single community.
This is due to the heuristics of these algorithms, 
where they lose granular structures and 
fail to identify communities for certain kinds of networks.
However, our approach keeps granular 
communities by focusing on the similarity of nodes
even when it has many dissimilar neighbors
but only a few similar ones.
It does not force small communities to join
to a giant component.
Our algorithm performs successfully 
on generated networks with planted 
community structure, 
i.e.,
ground-truth is known.
However, on real-life networks 
where ground-truth is created by using some meta-data,
all of the algorithms in benchmark find different results.
This may be due to the fact that 
meta-data does not reflect the actual ground-truth 
communities or meta-data shows 
different aspects of the network structure as 
discussed in the work of Peel et al.~\cite{Peel2017}.

With its local approach, \GCN\ is scalable
and suitable for distributed and parallel processing
(\emph{we have not 
implemented a parallel version for this paper}).
Community detection task can be split into separate subtasks
on many computation devices (with the necessary piece of network data), 
which will enable
real-time community detection on very large networks.\\

The source code of the algorithm is available at:
https://github.com/murselTasginBoun/CDBN

% =======================================
\section*{Acknowledgments}

Thanks to 
Mark Newman, 
Vincent Blondel and 
Martin Rosvall 
for the source codes of their community detection algorithms. 
Thanks to 
Mark Newman, 
Jure Leskovec and 
Vladimir Batagelj 
for the network datasets.

This work was partially supported 
by the Turkish State Planning Organization (DPT) TAM Project (2007K120610).

% =======================================
\bibliography{CDwBN}{}

% =======================================
\end{document}